\documentclass[reprint,superscriptaddress,amsmath,amssymb,aps]{revtex4-1}
\usepackage[utf8]{inputenc}
\usepackage[T1]{fontenc} 
\usepackage{graphicx}
\usepackage{rotating}
\usepackage{makeidx}
\usepackage{threeparttable}
\usepackage{float}
\usepackage{amsmath}
\usepackage{rotating}
\usepackage{hyperref}

\usepackage{relsize}
\usepackage{etoolbox}
\usepackage[most]{tcolorbox}
\usepackage{hyperref}
\usepackage{dcolumn}
\usepackage{bm}
\usepackage{wrapfig}
\usepackage{float}  
\usepackage[toc,page]{appendix}
\usepackage{gensymb}
\usepackage{epstopdf}
\usepackage{subfiles}
\usepackage{comment} 
\usepackage[autostyle=true]{csquotes} 
\usepackage{makecell}
\usepackage{siunitx}
\usepackage[all]{hypcap} 
\usepackage[most]{tcolorbox}
\usepackage{hyperref}
\usepackage{xcolor}
\usepackage{nicefrac}
\usepackage{chemformula}
\usepackage{ulem}  
\definecolor{gainsboro}{rgb}{0.86, 0.86, 0.86}
\usepackage[backgroundcolor=gainsboro]{todonotes}
\hypersetup{
  colorlinks   = true, 
  urlcolor     = blue, 
  linkcolor    = blue, 
  citecolor   = blue 
}

\newcommand{\weg}[1]{}

\newcommand{\obyth}[2]{$\frac{1}{3}$}
\newcommand{\obyf}[2]{$\frac{1}{4}$}

\newcommand{\orbs}{  $ d_{x^2-y^2}, d_{3z^2-r^2} $  }

\begin{document}

\title{Characteristic THz-emissions induced by optically excited collective orbital modes}

\author{Sangeeta Rajpurohit}
\email{srajpurohit@lbl.gov}
\affiliation{Molecular Foundry, Lawrence Berkeley National Laboratory, USA}

\author{Christian Jooss}
\affiliation{Institute for Material Physics, Georg-August-Universität Göttingen, Germany}

\author{Simone Techert}
\affiliation{Deutsches Elektronen-Synchrotron DESY, Notkestr. 85, 22607 Hamburg, Germany}
\affiliation{Institute of X-ray Physics, Georg-August-Universität Göttingen, Germany}

\author{Tadashi Ogitsu}
\affiliation{Lawrence Livermore National Laboratory, Livermore, USA}

\author{P.E. Blöchl}
\affiliation{Institute for Theoretical Physics, Clausthal University of Technology, Germany}
\affiliation{Institute for Material Physics, Georg-August-Universität Göttingen, Germany}

\author{L.Z. Tan}
\affiliation{Molecular Foundry, Lawrence Berkeley National Laboratory, USA}

\date{\today}

\begin{abstract}

We study the generation of collective orbital modes, their evolution, and the
characteristic nonlinear optical response induced by them in a photoinduced
orbital-ordered correlated oxide using real-time simulations based
on an interacting multiband tight-binding (TB) model. The d-d optical
transitions under femtoseconds light-pulse in an orbital-ordered state excite
collective orbital modes, also known as "orbitons". Consistently incorporating
electronic interactions and the interplay between charge, spin, and lattice
degrees of freedom in the TB-model provides a clearer understanding of how these
factors influence the generation and evolution of collective orbital modes. The dynamics of
Jahn-Teller vibrational modes in the photoinduced state modify the intersite orbital
interaction, which further amplifies these orbital modes.  In the presence of weak
ferroelectricity, the excitation of collective orbital modes induces a strong 
THz oscillatory photocurrent, which is long-lived. This suggests an alternative way to
experimentally detect low-energy collective modes through THz-emission
 studies in the photoinduced state. Our study also elucidates that quasiparticle dynamics in
improper ferroelectric oxides can be exploited to achieve highly interesting and
non-trivial optoelectronic properties.
\end{abstract}
\maketitle

\section {Introduction}
3d transition-metal oxides (TMOs) with strongly localized 3d orbitals due
to large onsite Coloumb repulsion belong to the strongly correlated material
class. The strong interaction among electron, spin, and
lattice degrees of freedom (DOF) in these oxides results in novel physical phenomena,
such as high-Tc superconductivity in layered cuprates, metal-insulator
transitions (Mott insulators) in nickel oxides, and colossal
magnetoresistance (CMR) and multiferroicity in manganites.
The low-lying excitations in these ordered states are
collective modes involving these DOFs such as phonons, magnons, and
orbitons \cite{Reticcioli2019,Essenberger2011,Fischer2009,Brink2001,Saitoh2001}.
The knowledge of the nature of the quasiparticles and collective
modes is important to understand the electrical and thermal transport properties
of correlated TMOs. In several TMOs, such as manganites, the splitting
of d-orbitals due to electronic interactions and the Jahn-Teller (JT) effect \cite{Jahn1937}
results in the formation of orbital-ordered states. The interactions between
orbitals at different sites in such materials allow propagation
of local orbital excitation to the entire lattice, resulting in
the formation of "orbitons". Similarly to collective spin
excitations in spin-ordered states, orbitons are elementary
excitations in materials that host a long-range orbital-order
pattern in the ground state \cite{Saitoh2001,Brink2000}.

Experimental observations of orbitons in materials remain challenging
due to the non-negligible coupling between orbital, phonon, and
spin degrees of freedom. Although orbiton excitation in orthorhombic
manganites, such as LaMnO$_3$, has been reported in a previous Raman
scattering study \cite{Saitoh2001}, the interpretation of several
low-energy peaks as orbitons by the authors has not been supported
by subsequent experiments. For example, resonating-inelastic X-ray
scattering experiments could not capture the orbiton in the energy range
suggested by earlier studies \cite{Inami2003,Tanaka2004}. Ultrafast
pump-probe experiments are powerful alternative tools for
studying quasiparticle excitations and their dynamics by selectively
modulating electronic, lattice, and spin degrees with optical excitations.
Polli et al. reported coherent orbital waves with
$\sim 31$ THz frequency \cite{Polli2007} in optically excited
$\rm{Pr_{1-x}Ca_xMnO_3}$.. The complex interaction between orbitons
and phonons in manganites is expected to create hybridized orbiton-phonon
excitations instead of pure orbitons and phonons \cite{Allen1999,Brink2001,Brink2000}.
The similar energy scales of the orbital super-exchange
interaction and electron-phonon coupling in manganites complicate the
interpretation of experimental quasiparticle peaks. Several previous theoretical
studies have demonstrated the crucial role of orbiton-phonon coupling
in controlling orbiton dispersion and frequency renormalization
\cite{Allen1999,Brink2001,Brink2000,Schmidt2007}. 
\begin{figure}[thp!]
     \begin{center}
     \includegraphics[width=\linewidth]{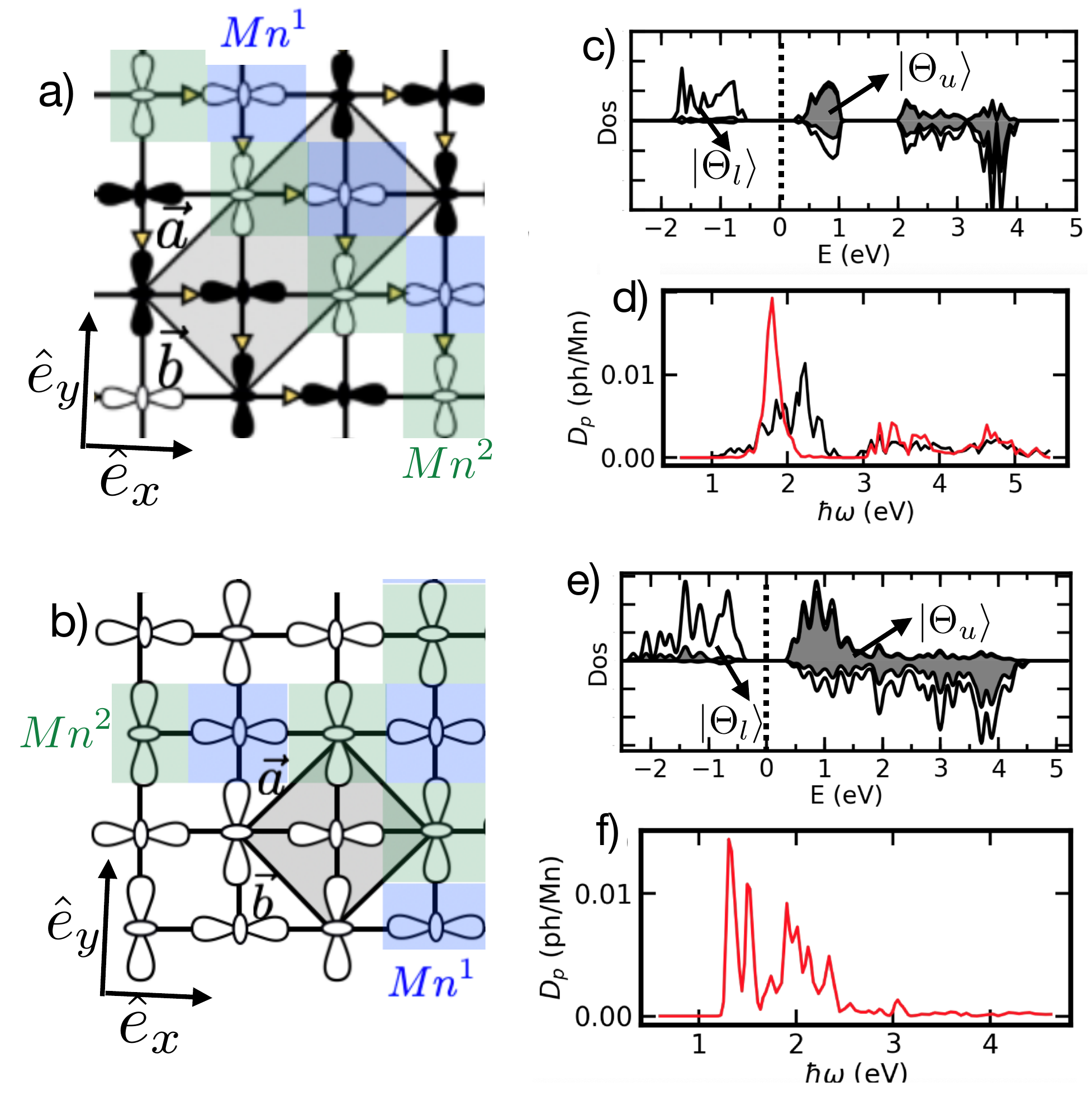}
     \caption{ a-b): SO and OO in E-type (a) and A-type (b) RMnO$_3$. 
     The black and white Mn-sites indicate spin-up  and spin-down
     Mn$^{3+}$. Spin-up Mn sites exhibiting orbital
     ordering along the x and y directions are shaded in blue and green, respectively. 
     Density of states of the calculated E-type (c) order projected onto 
     local $|\Theta_l\rangle$- and $|\Theta_u\rangle$-orbitals.
     The vertical dashed line indicates Fermi-level. Photon absorption
     density $D_p$ per Mn of the E-type (d) order under 30 fs light-pulse
      as a function of photon energy
      $\hbar\omega$. (e-f): Corresponding density of states (e) and absorption
      (f) plots of the A-type order. The red and black curves indicate $D_p$ under
      light-pulse polarized along direction $\vec{a}=\hat{e}_x-\hat{e}_y$ and
      $\vec{b}=\hat{e}_x+\hat{e}_y$,  respectively. }
     \label{fig:fig1}
     \end{center}
\end{figure}

In the present work, we theoretically study the nonlinear optical
response of orbitons, and show that it can serve as a method for their experimental
detection. We model the evolution of orbitons using real-time simulations based on an
interacting three-dimensional tight-binding (TB) model.  The TB model
explicitly takes into account interacting electrons and their coupling to spin and
phonons, which is needed to accurately describe the collective modes of these
active degrees of freedom. We report low-lying collective orbital excitations
by an optical pump in orbital-ordered states, which exhibit symmetry breaking
in the orbital-space. In these states, the orbital super-exchange mechanism induced
by the restrictive intersite hybridization gives rise to a non-local orbital
interaction and allows propagation of the local orbital excitations as a collective mode. 
The reported collective orbital modes have
characteristic frequencies of $\sim5.0$ and $90$ THz. 

Interestingly, in the presence of weak ferroelectricity,
the collective orbital modes induce a non-linear optical response in the form
of THz emission. We attribute this to the formation of polar
orbitons induced by the breaking of centrosymmetry in the ferroelectric phase,
resulting in symmetry-allowed directional THz current.

\section {Tight-binding model}
The spin- and orbital-order (OO) of the two orthorhombic manganites RMnO$_3$
considered in this study, A-type and E-type, are shown in Figure \ref{fig:fig1} (a-b).
In the A-type spin order (SO), the $ab$-planes are ferromagnetic. On the
other hand, the $ab$-plane in the E-type SO consists of quasi-one-dimensional
ferromagnetic zigzag chains that are antiferromagnetically aligned with each other.
The adjacent ab-planes in both SOs are stacked antiferromagnetically on top of each
other in the z-direction.

To study orbital excitations, their evolution and decay in the A- and E-type
magnetic orders of RMnO$_3$, we combined a 3d TB-model with real-time simulations.
The model considers the octahedral crystal-field splitting of the Mn 3d-shell
into three non-bonding $t_{2g}$-orbitals and two anti-bonding $\rm{e_g}$-orbitals.
The e$_g$ orbitals are explicitly incorporated in the model, whereas the localized
$t_{2g}$-electrons are treated as classical spins $\vec{S}_R$. In addition to
$\rm{e_g}$-electrons and $\rm{t_{2g}}$-spins, we also take into account lattice
degrees of freedom by considering the local breathing mode Q$_{1,R}$ and Jahn-Teller
(JT) active modes Q$_{2,R}$ and Q$_{3,R}$ oxygen octahedra.
The potential energy of the system is expressed as 
\begin{eqnarray}
E_{pot}\Big(|\psi_n\rangle,\vec{S}_{R},Q_{i,R}\Big)
&=&E_{e}(|\psi_n\rangle) +E_{S}(\vec{S}_R)+E_{ph}(Q_{i,R})\nonumber\\
&&\hspace{-2cm}+
E_{e-ph}(|\psi_n\rangle,Q_{i,R})+E_{e-S}(|\psi_n\rangle,\vec{S}_R)
 \label{eq:tbm}
\end{eqnarray} 
in terms of one-particle states 
\begin{eqnarray}
|\psi_n\rangle{=}\sum\limits_{\sigma,\alpha,i} |\chi_{\sigma,\alpha,i}\rangle \psi_{\sigma,\alpha,i,n}
\label{eq:wf}
\end{eqnarray} 
of e$_g$-electrons, $t_{2g}$-spin $\vec{S}_R$ and oxygen octahedral  phonon modes $Q_{i,R}$.
The basis set $|\chi_{\sigma,\alpha,i}\rangle$'s for the one-particle states
consists of local spin orbitals with spin $\sigma\in\{\uparrow,\downarrow\}$ and
orbital character $\alpha\in$\orbs. The one-particle states are then defined by the coefficients $\psi_{\sigma,\alpha,i,n}$.

The kinetic energy $E_{hopp}$ of $e_g$-electrons and the
electron-electron interaction $E_{coul}$ contribute to the energy $E_{e}=E_{hopp}+E_{coul}$
of the electronic system. The eg-electrons delocalize between Ni sites
through intermediate oxygen bridges, and the kinetic energy
term $E_{hop}$ is expressed as 
\begin{eqnarray}
E_{hop}&=&\sum\limits_{R,R',\sigma,n}
 f_n \sum\limits_{\alpha,\alpha'} \psi_{\sigma,\alpha,R,n}T_{\alpha,\alpha',R,R'}\psi^*_{\sigma,\alpha',R,n}
\label{eq:tbm_2}
\end{eqnarray} 
where $T_{\alpha,\alpha',R,R'}$ is the hopping matrix element;
see the appendix. The oxygen contribution is down-folded and built into
the Ni-d orbitals. The hopping-matrix elements contribute only
onsite and nearest-neighbor terms between the Mn sites. The hopping
matrix elements along $x$, $y$ and $z$ 
directions are defined as 
\begin{eqnarray}
T^{x/y}&=&-t_{hop}\left(\begin{array}{cc}
  3/4   & \mp\sqrt{3}/4 \\
  \mp\sqrt{3}/4   & 1/4
\end{array}
\right)
\nonumber\\
T^{z}&=&-t_{hop}\left(\begin{array}{cc}
  0   & 0 \\
  0  & 1
\end{array}
\right)
\end{eqnarray}
The onsite Coulomb energy between $e_g$ electrons
\begin{eqnarray}
\label{eq:E-hartree}
E_{coul}&=&
\frac{1}{2}(U-3J_{xc})\sum_{R}
\left(\sum_{\sigma,\alpha}
\rho_{\sigma,\alpha,\sigma,\alpha,R}\right)^2 \nonumber \\ &-&\frac{1}{2}(U-3J_{xc})\sum_R
\sum_{\sigma,\alpha,\sigma',\beta}
|\rho_{\sigma,\alpha,\sigma',\beta,R}|^2 \nonumber \\
&+&\frac{1}{2}J_{xc}\sum_R\sum_{\sigma,\sigma'}(-1)^{\sigma-\sigma'}\sum_{k\in\{x,z\}}
\nonumber\\
&\times&\Bigl[
\Bigl(\sum_{\alpha,\beta}\rho_{\sigma,\alpha,\sigma',\beta,R}
\sigma^{(k)}_{\beta\alpha}\Bigr)
\Bigl(\sum_{\alpha,\beta}\rho_{-\sigma,\alpha,-\sigma',\beta,R}
\sigma^{(k)}_{\beta\alpha}\Bigr)
\nonumber\\
&&
+\Bigl(\sum_{\alpha}\rho_{\sigma,\alpha,\sigma',\alpha,R}\Bigr)
\Bigl(\sum_{\alpha}\rho_{-\sigma,\alpha,-\sigma',\alpha,R}\Bigr)\Bigr]
\end{eqnarray}

The first and second terms in the expression $E_{coul}$ are
the Hartree and the corresponding self-interaction
correction. Here, $\rho_{\sigma,\alpha,\sigma',\beta,R}$
are the local one-center reduced density matrix.
$\sigma^{(k)}_{\alpha,\beta}$ indicates 
the three Pauli matrices for $k\in\{x,y,z\}$.
The notation $-\sigma$ implies $-\sigma=\uparrow$
for $\sigma=\downarrow$ and vice versa. Similarly,
$(-1)^{\sigma-\sigma'}=1$ for $\sigma=\sigma'$ and
$(-1)^{\sigma-\sigma'}=-1$ for $\sigma\neq\sigma'$.
The elements of the reduced-density matrix
${\hat{\rho}}$ are defined as 
\begin{equation}
\rho_{\sigma,\alpha,R,\sigma',\alpha',R}{=}\sum\limits_{n}
f_n\psi_{\sigma,\alpha,R,n}\psi^*_{\sigma',\alpha',R,n}.
\end{equation}
Here $f_n$ are the occupations of the single-particle
states $|\psi_n\rangle$. The el-ph coupling $E_{e-ph}$ is
\begin{equation}
\label{eq:E-vib-04-08-2016}
E_{e-ph} = g_{JT} 
\sum_{R,\sigma}\sum_{\alpha,\beta} \rho_{\sigma,\alpha,\sigma,\beta,R}
M^Q_{\beta,\alpha}(Q_{1,R},Q_{2,R},Q_{3,R}).
\end{equation}
Here $g_{JT}$ and $g_{br}$ are the el-ph coupling constants
and $\mathbf{M}^Q(Q_{1,R},Q_{2,R},Q_{3,R})$ is defined as
\begin{eqnarray}
\mathbf{M}^Q(Q_{1,R},Q_{2,R},Q_{3,R})=
\left(\begin{array}{cc}
Q_{3,R} & Q_{2,R}\\Q_{2,R} & -Q_{3,R}
\end{array}\right)-{\bm 1}\frac{g_{br}}{g_{JT}}Q_{1,R}\;.
\end{eqnarray}
The energy of the phonon subsystem consists of restoring energy
\begin{eqnarray}
E_{ph}&=&\frac{1}{2}k_{JT}\sum_{R}\Big(Q^2_{2,R}+Q^2_{3,R}+\frac{k_{br}}{k_{JT}}Q^2_{1,R}\Big).
\label{eq:tbm_3}
\end{eqnarray} 

The term $E_{S}(\vec{S}_R)$ is Heisenberg-like intersite
antiferromagnetic coupling between $t_{2g}$-spins $\vec{S}_R$ and defined as

\begin{equation}
\label{eq:E-vib-04-08-2016}
E_{S} = J_{AF}
\sum_{<R,R'>} \big(\frac{3\hbar}{2}\big)^{-2}\vec{S}_R\vec{S}_R'.
\end{equation}
Here, $<RR'>$ indicates nearest-neighbor Mn sites.

The Hund's coupling term defined as $t_{2g}$-spin $\vec{S}_R$.
\begin{equation}
\label{eq:E-vib-04-08-2016}
E_{e-S} = J_H 
\sum_{R,\alpha}\sum_{\sigma,\sigma'} \rho_{\sigma',\alpha,\sigma,\alpha,R}
M^S_{\sigma',\sigma}(\vec{S}_R).
\end{equation}
where matrix $M^S(S_R)$ is defined as
\begin{eqnarray}
\mathbf{M}^S(S_R)=
\left(\begin{array}{cc}
S_z & S_x+iS_y \\S_x-iS_y & -S_z
\end{array}\right)-{\bm 1}.
\end{eqnarray}
which tends to align the spin of $e_g$-electron along
local $t_{2g}$-spin direction. For complete details of the
model and its parameters, we refer to \cite{Sotoudeh2017,Rajpurohit2020,Rajpurohit2020_2}.
The model parameters used in the present work are
summarized in Table I.

\begin{table}[!ht]
\begin{center}
\begin{tabular}{|lrl|lrl|}
\hline
\hline
  $J_H$          & 0.653&eV 
& $g_{br}$       & 2.988 &eV/\AA \\
  $U$            & 2.514 &eV 
& $k_{br}$       & 12.04 &eV/\AA$^2$ \\
  $J_{xc}$       & 0.692 &eV
& $J_{AF}$       & 0.014  &eV \\
  $g_{JT}$           & 2.113  &eV/\AA 
& $t_{hop}$ & 0.585  & eV \\
  $k_{JT}$         & 5.173  &eV/\AA$^2$ 
& $\bar{d}$& 1.923&\AA \\
\hline
\end{tabular}
\caption{\label{tab:parameters} Parameters for the
TB-model. For the detailed information about
the extraction of these model parameters from the
first-principle studies, see reference \cite{Sotoudeh2017}.}
\end{center}
\end{table}

Unlike the previous simple 1d model-based studies of orbitons, the above
TB-model takes into account electron-phonon, electron-electron, and
electron-spin interaction on an equal footing, allowing for a systemic
investigation of their combined effect on orbital physics in manganites.
In perovskites manganites, an octahedral distortion around one Mn site is coupled
to distortions around neighboring Mn sites, resulting in a cooperative lattice effect,
allowing propagation of interaction between eg-orbitals at different Mn sites. The
model takes into account this cooperative nature of the octahedra distortion.

\subsection{Ground-state }
Figures \ref{fig:fig1}-c and \ref{fig:fig1}-e show the density of states in the ground state
of the E- and A-types $\rm{RMnO_3}$, respectively, calculated using the above TB model
(Eq. \ref{eq:tbm}) by optimizing the electronic and structural degrees
of freedom while keeping the $\rm{t_{2g}}$-spin configuration fixed.
In the ground state, all Mn-sites have the oxidation state ${3+}$ and
are JT active. The JT-effect at Mn$^{3+}$ site lifts the degeneracy of
the local $\rm{e_g}$-orbitals. The lower filled $\rm{e_g}$-states $|\Theta_{l,R}\rangle$
in the ground state are described by the linear combination
\begin{eqnarray}
|\Theta_{l,R}\rangle = -\sin{(\gamma)}|d_{x^2-y^2}\rangle \pm \cos{(\gamma)}|d_{3z^2-r^2}\rangle
\label{eq:eg_1}
\end{eqnarray} 
of the $\rm{e_g}$-orbitals $d_{x^2-y^2}$ and $d_{3z^2-r^2}$. Here $R$ is
the site index and $\gamma{=}45^{\circ}$. 
The corresponding unoccupied states $|\Theta_u\rangle_R$ are 
\begin{eqnarray}
|\Theta_{u,_R}\rangle = -\sin{(\gamma)}|d_{x^2-y^2}\rangle \mp \cos{(\gamma)}|d_{3z^2-r^2}\rangle.
\label{eq:eg_2}
\end{eqnarray} 
The upper sign in front of the $|d_{3z^2-r^2}\rangle$-state
describes an orbital polarization along $x$ and the lower sign describes
an orbital polarization along the $y$ direction.

Our model predicts both the A-type and E-type magnetic states as band-insulator
with band-gap of 1.02 eV and 0.89 eV, respectively. This band gap arises
predominantly from the JT-splitting $\Delta_{JT}=2g_{JT}\sqrt{Q_1^2+Q_2^2}$
of the majority-spin states at Mn$^{3+}$ sites and is sensitive
to the amplitude of the JT modes $Q_i$.

The local $\rm{e_g}$-orbital polarizations
form a long-range OO in the ground-state as shown in Figure \ref{fig:fig1} a-b,
where the orbital polarization alternates between the Mn sites along the $x$- and $y$-directions. 
In both A-type and E-type magnetic states, the lower eg-orbital $|\Theta_{l,R}\rangle$ at every
$R^{th}$ Mn-site hybridizes with the upper $|\Theta_{u,R+m}\rangle$ orbital
located at the nearby spin-aligned $(R+m)^{th}$ site pointing towards the former. 
The E-type state is improper ferroelectric where
ferroelectricity originates from the restricted Mn-Mn exchange
interaction between spin-aligned Mn sites \cite{Mochizuki2011}.
The charge-center between the spin-aligned Mn sites moves off the
center, indicated by a yellow arrow in Figure \ref{fig:fig1}-a. 
Our model calculations predict a net polarization $\sim 18.2$ $\rm { \mu C/cm^2}$
in the $\vec{b}$ direction in the E-type state (more information in SI). 
This bulk polarization value is slightly higher than the theoretical value predicted
from First-principles based studies \cite{Picozzi2008}. The current model
includes only the valence eg-electrons and does not consider the polarization contribution
from the core electrons, A-type cations, and oxygen anions.

\begin{figure}[thp!]
     \begin{center}
     \includegraphics[width=1.0\linewidth]{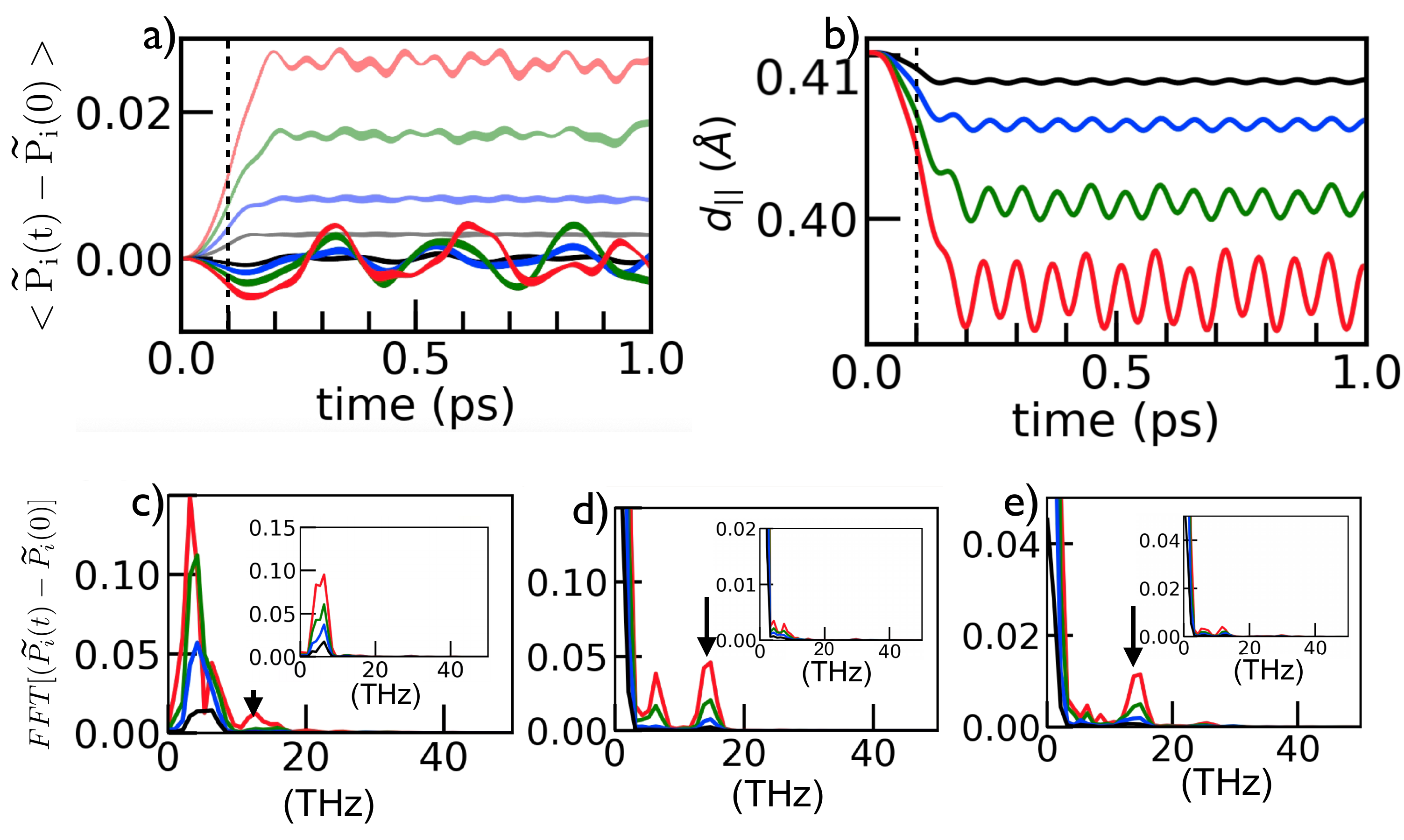}
     \caption{ Dynamics of the on-site orbital-polarization and phonon modes in the E-type. 
     a-b: Evolution of the instantaneous orbital-polarization
     vector $\langle\vec{P}_R\rangle$ (a) and changes $\Delta d_{||}$ of longer O-O
     bonds of the $ab$-plane (b) at different light intensities. 
    $\langle P_{R,y}\rangle$ and $\langle P_{R,z}\rangle$ components are indicated by
    solid and partially transparent lines. are plotted in the main figure, while the
    inset shows $\langle \vec{P}_R,y\rangle$. c-e: Fourier transform of the
    components $x$ (d), $y$ (c) and $z$ (e) of $\vec{P}_R(t)$ in the presence (main)
    and absence (inset) of atom dynamics. The intensities are color-coded as described
    in Figure 2. Colors indicate the strength of the light field: $A_o=0.04$ $\hbar/ea_o$) (black), $A_o=0.06$ $\hbar/ea_o$)
    (blue), $A_o=0.08$ $\hbar/ea_o$) (red), $A_o=0.10$ $\hbar/ea_o$) (green).
    The dashed vertical line is the center of 100-fs Gaussian light-pulse.}
     \label{fig:fig2}
     \end{center}
\end{figure}

\section{Real-time simulations of photo-excitation and its evolution}
We demonstrate the generation of collective orbital modes in the photoinduced
A-type and E-type states using real-time time-dependent density-functional
theory formalism based on the TB-model defined in Eq.\ref{eq:tbm}. 
The one-particle electron wavefunctions evolve under the time-dependent Schrödinger
equation
\begin{eqnarray}
i\hbar\partial_t \psi_{\sigma,\alpha,R,n}&=&
\frac{\partial E_{pot}}{\partial\psi^*_{\sigma,\alpha,R,n}}
\end{eqnarray}
In our simulations, the effect of phonons is simulated using Ehrenfest dynamics. 
The oxygen atoms obey Newton's equations of motion.
\begin{eqnarray}
M_O\partial^2_t R_j&=& -\frac{\partial E_{pot}}{\partial R_j}
\end{eqnarray}
$R_j$ are the structural degrees of freedom of the oxygen ions and
$M_O$ is their mass. 

The t$_{2g}$-spins are kept fixed during the simulations. The effect
of the electric-field defined by the vector potential
$\vec{A}(t)=\vec{e}_s \omega \rm{Im}(A_oe^{-i\omega t})g(t)$, 
where $A_o$, $\omega$, and $\vec{e}_s$ is the amplitude of
the vector potential, angular frequency, and direction of the
electric-field, implemented in our model through the Peierls
substitution method \cite{Peierls1933}. We consider a Gaussian-shape
of the pulse imposed by $g(t)$. We study the photocurrent dynamics under
a finite Gaussian-shaped light-pulse using a 8×8×2 super-cell with
512 atoms, 128 Mn, and 384 oxygen atoms. The simulations are performed
with periodic boundary conditions with k-points sampling at
$\Gamma$ and temperature $T=0$. Table II summarizes the relevant
parameters we used for the simulations.
\begin{table}[!htb]
\centering 
\label{tab:t1}
\begin{tabular}{|l|l|}
\hline\hline
k-grid  & $1{\times}1{\times}1$ \\
supercell  & $N_x{\times}N_y{\times}N_z{=}8{\times}8{\times}2$\\ 
Mn sites per unit cell  &$N_{Mn}=128$\\
O sites per unit cell  & $N_{O}=384$\\
lattice vectors  & $\vec{a}=\vec{b}=5.628$ ~\AA, $\vec{c}=7.60$ ~\AA\\
Mn-Mn spacing & $d_{Mn-Mn}=3.84$~\AA\\ 
time step  & $\Delta_t{=}0.060 (4\pi\epsilon_0)^2\hbar^3/(m_ee^4)$\\
          & $= 1.45 \times 10^{-18} $~s\\
oxygen mass   &  $M_{O}=15.998$~u  \\
pulse length (FWHM) & $2\sqrt{\ln2}c_w=100~{\rm fs}$ \\

\hline
\hline
\end{tabular}
\caption{Simulation parameters.}
\end{table}  

\subsection{Optical absorption}

Figure \ref{fig:fig2} (g-h) shown the spectral-distribution of optical absorption,
obtained by calculating the photon-absorption density $D_p$ (total number of photons
absorbed per site). We compute $D_p{=}\delta E_{pot}/{\hbar\omega}$ from the total change in energy,
defined in Eq. \ref{eq:tbm}, before and after a 30-fs Gaussian-shaped light pulse.
The A-type (E-type) state exhibits a broad absorption peak around
frequency $\omega_{p_1}{=}1.30$ (2.0) eV. We assign this peak to dipole-allowed intra-chain
electronic transitions from the majority-spin $|w_1\rangle$-states,
which are bonding states of lower $|\Theta_l\rangle$ and upper
$|\Theta_u\rangle$ orbitals pointing toward each other and are located
at spin-aligned neighboring sites, to the corresponding anti-bonding
states $|w_2\rangle$. We select $\omega_{p_1}{=}1.33$ ($\omega_{p_1}{=}1.79$)
to simulate the evolution of optical excitations into collective orbital modes
in the A-type (E-type) state under a 100-fs light pulse with light-polarization
in the $ab$-plane along $\vec{a}$ ($\vec{a}$ and $\vec{b}$). The  
results remain qualitatively similar for the light polarization along 
$\vec{a}$ and $\vec{b}$ in the E-type state and we discuss only the $\vec{b}$ case here. 

\begin{figure}[tp!]
     \begin{center}
     \includegraphics[width=\linewidth]{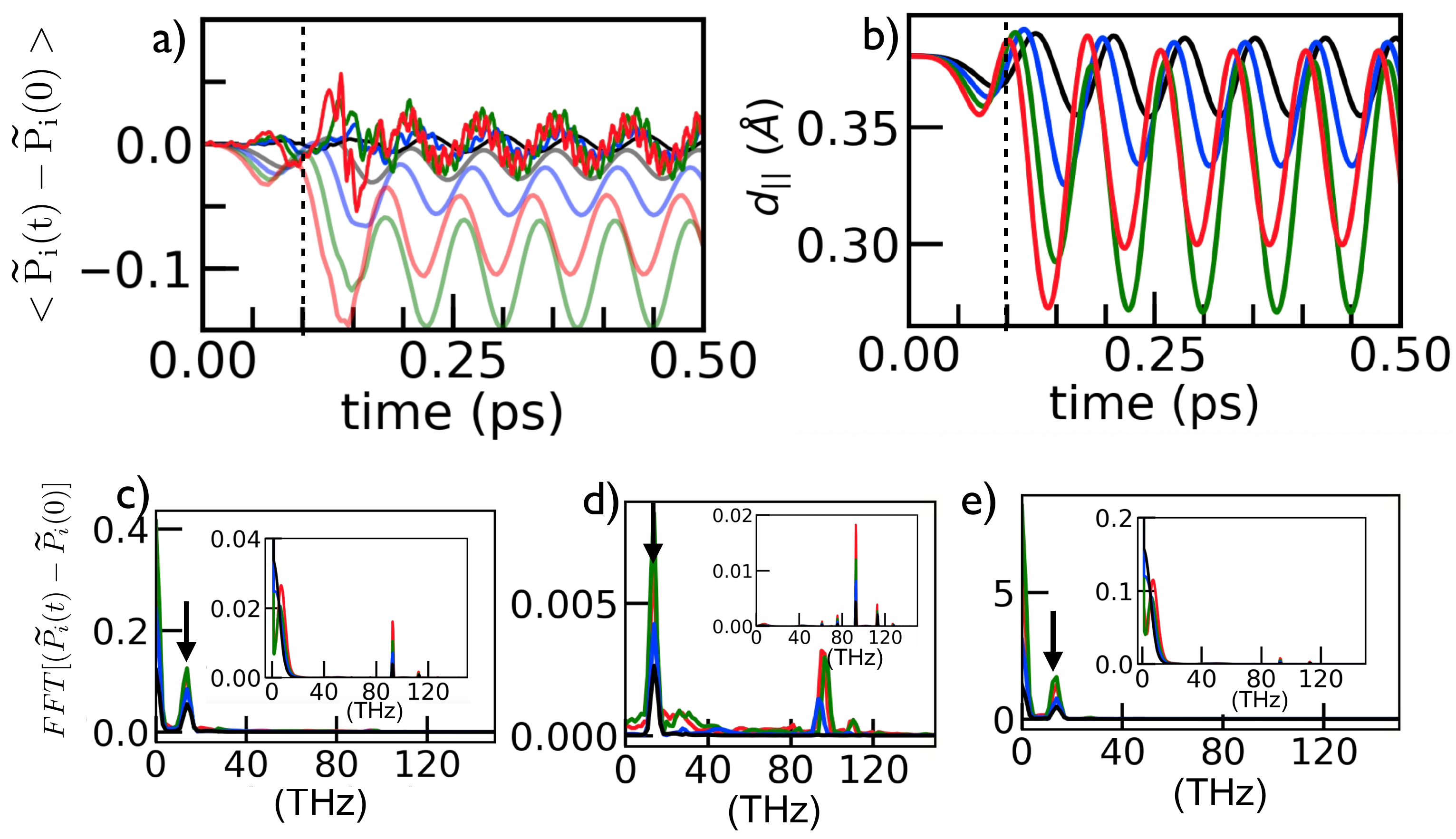}
     \caption{Dynamics of the on-site orbital-polarization and phonon modes in the A-type.
     a-e: as labeled in Figure \ref{fig:fig1}. Colors indicate the strength of the light field: $A_o=0.15$ $\hbar/ea_o$) (black), $A_o=0.20$ $\hbar/ea_o$)
    (blue), $A_o=0.25$ $\hbar/ea_o$) (red), $A_o=0.30$ $\hbar/ea_o$) (green).}
     \label{fig:fig3}
     \end{center}
\end{figure}

Figures \ref{fig:fig2} (a-b) and \ref{fig:fig3} (a-b) show the dynamics
of local orbital polarization and phonon modes during
and after the 100-fs light-pulse with different intensities ranging
from $A_o=0.05$ to $0.30$ $\hbar/ea_o$.  To follow
the changes in local orbital polarization in our simulations, we
define an orbital polarization vector $\vec{\tilde{P}}_R(t)$
\begin{eqnarray}
\tilde{P}_{R,x}(t)&=&\sum_{\sigma} \tilde{\rho}_{\sigma,\tilde{\alpha},R,\sigma,\beta,R}(t)-\tilde{\rho}_{\sigma,\tilde{\beta},R,\sigma,\tilde{\alpha},R}(t)\nonumber \\
\tilde{P}_{R,y}(t)&=&\sum_{\sigma}-i\tilde{\rho}_{\sigma,\tilde{\alpha},R,\sigma,\tilde{\beta},R} (t)+i\tilde{\rho}_{\sigma,\tilde{\beta},R,\sigma,\tilde{\alpha},R}(t) \nonumber \\
\tilde{P}_{R,z}(t)&=&\sum_{\sigma}\tilde{\rho}_{\sigma,\tilde{\alpha},R,\sigma,\tilde{\alpha},R}(t)-\tilde{\rho}_{\sigma,\tilde{\beta},R,\sigma,\tilde{\beta},R}(t)
\end{eqnarray}
for site R. Here $\hat{\tilde{\rho}}(t)$ is the instantaneous on-site one-particle
reduced density matrix at time $t$ in the basis set of local orbitals
 $\{|\Theta_{l,R}\rangle, |\Theta_{u,R}\rangle\}$ calculated using the expression 
\begin{eqnarray}
\tilde{\rho}_{\sigma,\tilde{\alpha},R,\sigma,\tilde{\beta},R}(t)&=&\sum_{n} f_n \sum_{\sigma}
\langle \psi_{n}(t)|\Theta_{\tilde{\alpha},\sigma,R} \rangle \nonumber \\ 
&& \langle \Theta_{\tilde{\beta},\sigma,R}|\psi_{n}(t)\rangle. 
\label{eq:dens_mat}	 		 
\end{eqnarray}for every site $R$. Here indices $\tilde{\alpha}$ and $\tilde{\beta} \in \{|\Theta_l\rangle,|\Theta_u\rangle\}$ (where $|\Theta_l\rangle$ and $|\Theta_u\rangle$
are defined in Eq. \ref{eq:eg_1} and \ref{eq:eg_2}) and $f_n$ are the occupancies of 
the one-particle wavefunctions $|\psi_{n}(t)\rangle$. The occupancies $f_n$ are fixed, $f_n=1$
for the occupied states, and $f_n=0$ for the unoccupied states.

The calculated on-site polarization vector $\vec{\tilde{P}}_R$ (t) has values $(-0.18,0.00,-0.75)$
and $(0.11,0.00,0.70)$ in the E-type and A-type ground-state, respectively. 
The electronic transitions during photoexcitation transfer electrons from the
occupied $|\Theta_{l,R}\rangle$ state at site $R$ to the empty $|\Theta_{u,R+m}\rangle$ 
at nearby spin-aligned $(R+m)$th sites. This results in modification 
of the the local orbital polarization at Mn sites which is reflected in the changes in the
$\tilde{P}_{R,z}$ component; see Figures \ref{fig:fig2}-a and \ref{fig:fig3}-a. This modification 
in local orbital-polarization excites the octahedral JT-modes that are strongly
coupled to $\rm{e_g}$-orbital occupancies, see Figure \ref{fig:fig2}-b and \ref{fig:fig3}-b.
On photo-excitation the shorter (longer) O-O bond distances $d_{\perp}$ ($d_{||}$)
in the $ab$-plane start to increase (decrease). The oscillations of the JT-mode are
enhanced at higher light intensities. 

A careful analysis of the dynamics of the local $\vec{\tilde{P}}_{R}$ vectors
indicates oscillating components with frequencies other than JT modes.
The Fourier transform of $\vec{\tilde{P}}_{R} (t)$ shows multiple
peaks (see Figures \ref{fig:fig2} c-e and \ref{fig:fig3} c-e). Besides the strongest
peak close to the light-field frequency, both the E-type and the A-type exhibit a peak at
5.0 THz in the absence of atom dynamics. The A-type exhibits another strong
peak at $\sim 90$ THz. The lack of a broadband THz emission suggests
the presence of collective excitations at a set of characteristic natural frequencies.
We thus attribute these THz peaks to collective orbital modes non-resonantly
excited by the optical pump. In the presence of atom dynamics, a distinct
peak at $\sim 14$-THz appears in both the A-type and the E-type (indicated by
the black vertical arrow) due to the JT-mode dynamics.

\begin{figure}[thp!]
     \begin{center}
     \includegraphics[width=\linewidth]{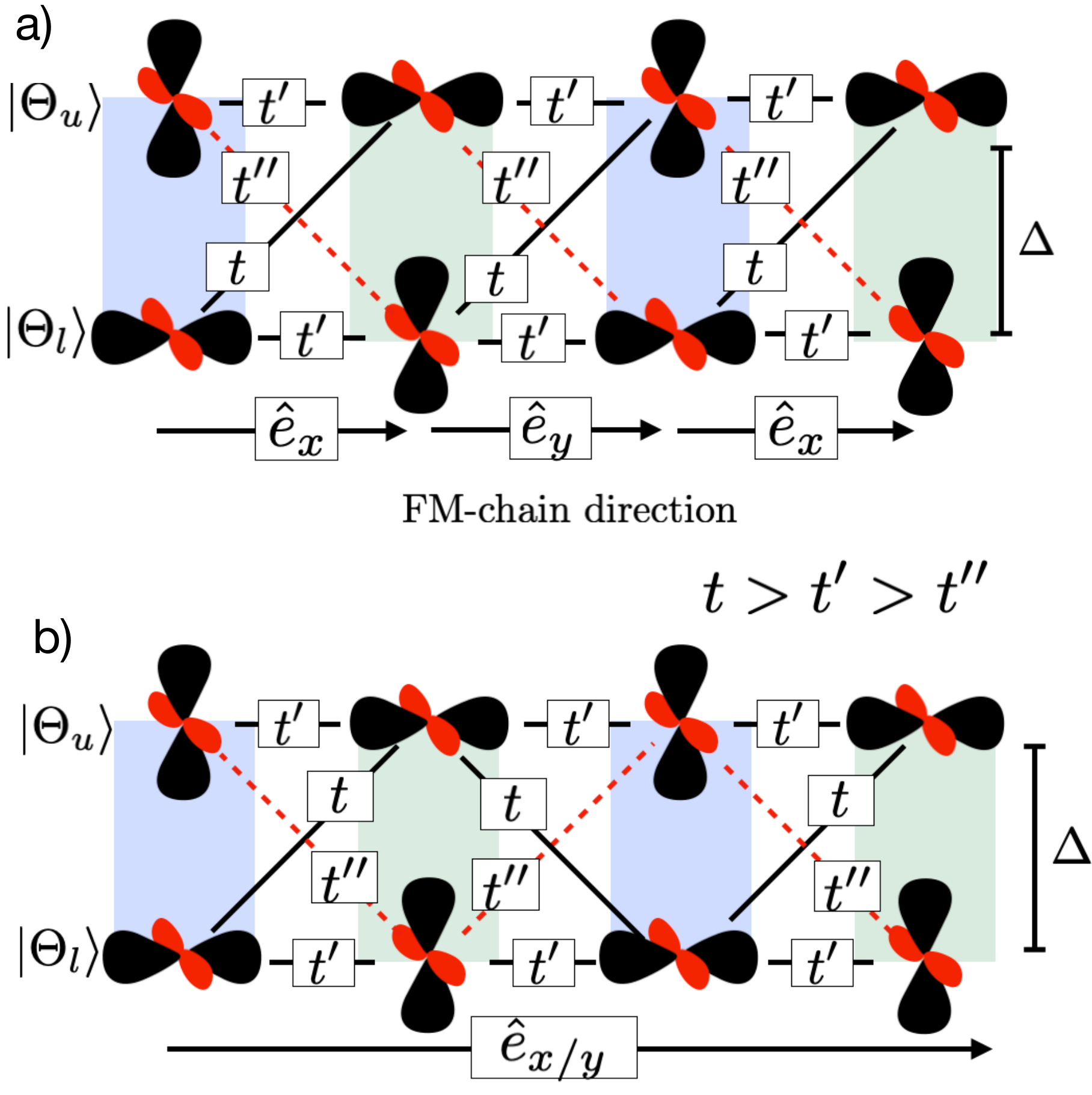}
     \caption{A simplified two-band model consisting of hoppings ($t,t',t'')$ between occupied
     $|\Theta_l\rangle$ and empty $|\Theta_l\rangle$ orbitals
     along FM zig-zag chain direction in E-type (a) and the x or y direction
     in the A-type (b). The $\hat{e}_x$ and $\hat{e}_x$ bond vectors
     are defined in Fig.~\ref{fig:fig1}.The on-site splitting $\Delta=\Delta_{JT}+U-3J$ due to
     Jahn-teller effect and Coloumb interaction at the Mn site forms staggered on-site potentials for eg-orbitals.}
     \label{fig:fig6}
     \end{center}
\end{figure}

The low-energy spectrum of the orbital-waves in A-type and E-type
states can be understood from the effective orbital super-exchange interactions
originating from the hybridization between orbital polarized sites. To determine
the effective orbital super-exchange interaction, we consider a simplified two-band model,
illustrated in Figure \ref{fig:fig6}, in $\{|\Theta_l\rangle,|\Theta_u\rangle\}$ basis. 
The model takes into account i) onsite energies $\sum_R\tilde{E}^{on-site}_R$ of the local orbitals
$\{|\Theta_{l,R}\rangle$ and $|\Theta_{u,R}\rangle\}$ and ii) the electron hopping contribution
$\sum_{<R,R'>}\tilde{E}^{hop}_{R,R'}$ with effective hopping elements $\tilde{t}$, $\tilde{t}'$
and $\tilde{t}''$ between nearby sites $R$ and $R'$.  
The energies of the local orbitals $\{|\Theta_l\rangle$ and $|\Theta_u\rangle\}$ in
this simplified model are obtained by considering the model defined in Eqn. \ref{eq:wf} in
the atomic limit. In this limit, the on-site energy contribution
$\tilde{E}^{on-site}=\tilde{E}_{coul}+\tilde{E}_{e-ph}+ \tilde{E}_{ph}$ includes the Coulomb energy 
$\tilde{E}_{coul}$, electron-phonon coupling $\tilde{E}_{e-ph}$ and the phononic contribution
 $\tilde{E}_{ph}$. \cite{Sotoudeh2017}. We calculate
the energy of the on-site orbitals using Janak's Theorem. This involves taking the
derivative of the on-site energy functional $\tilde{E}^{on-site}$ with respect to the occupations
$\tilde{n}_{l}$ and $\tilde{n}_{u}$ of the orbitals $|\Theta_l\rangle$ and $|\Theta_u\rangle$, 
respectively. We choose occupancies $\tilde{n}_{l}{=}1$ and $\tilde{n}_{u}{=}0$. Orbiton frequencies
would then correspond to the eigenfrequencies of this simplified two-band model.
For the FM chains in the E-type, the unit cell of the isolated zigzag chain has
two Mn sites with local $Q_i$ mode.  The values of effective hoppings $\tilde{t}$,
$\tilde{t}'$ and $\tilde{t}''$ are $-0.93t_{hop}$ $-0.25t_{hop}$ and $-0.067t_{hop}$,
respectively, see SI \cite{supp} for details (see also references \cite{Sotoudeh2017, Jahn1937, Kanamori1960} therein).
The weak orbital super-exchange interaction
$\tilde{J}^1 \sim \tilde{t}''^2/(\Delta_{JT}+U-3J)$, originating from the
hybridization of orthogonal $\rm{e_g}$-orbitals $|\Theta_{l,R}\rangle$ and
$|\Theta_{u,R+m}\rangle$ at the spin-aligned sites Mn$_R$ and Mn$_{R+m}$
dictates the frequency of the collective orbital modes in the E-type
phase. The frequency of collective orbital modes in the A-type depends on 
$\tilde{J}^1$ as well the relatively stronger orbital super-exchange interaction
$\tilde{J}^2\sim \tilde{t}^2/(\Delta_{JT}+U-3J),$ induced by the
hybridization $\tilde{t}$ between $\rm{e_g}$-orbitals with similar
orbital-polarization at nearby spin-aligned Mn-sites.

\subsection{THz-emission induced by collective orbital modes in ferroelectric E-type}
The excited collective orbital modes can modulate the optical conductivity
which in turn can induce a strong nonlinear optical response in the form of
low-frequency THz-emission in the presence of ferroelectricity.
To investigate the THz-emission of ferroelectric E-type RMnO$_3$, 
we calculate the photocurrent generation and its evolution under a 100-fs light pulse.
The instantaneous total photocurrent density $\vec{j}^{tot}(t)$,
is defined as 
\begin{eqnarray}
 \vec{j}^{tot}(t)&=&\sum_n f_n\sum\limits_{R' \epsilon \langle NN\rangle}  \sum_{\sigma}\sum_{\alpha,\beta}\cdot\Big(\psi^*_{\sigma,\alpha,R,n} (t) 
 T_{\alpha,\beta, R,R'}\psi_{\sigma,\beta,R',n}  (t) \nonumber\\
&-&\psi^*_{\sigma,\beta,R',n} (t)T_{\beta,\alpha,R',R}\psi_{\sigma,\alpha,R,n} (t)  \Big) \vec{e}_{R-R'},
   \label{eq:charge_current}
\end{eqnarray}
In the above equations,
$V{=}d_{Mn-Mn}^3N_R$ is the volume of
the unit cell with the total $N_R$ Mn-sites
and the average Mn-Mn bond length 
$d_{Mn-Mn}{=}\SI{3.845}{\angstrom}$.
Here, $f_n$ is the occupancy of the one-particle wavefunctions
$|\psi_n\rangle$, $\vec{e}_{R-R'}=\Big(\frac{\vec{R}-\vec{R'}}{|\vec{R}-\vec{R'}|}\Big)$
is the unit vector in the direction joining the sites $R$ and $R'$, and
$T'_{\alpha,\beta, R,R'}=T_{\alpha,\beta, R,R'}e^{-i\vec{A}(t)(\vec{R}-\vec{R'})}$,
where $T_{\alpha,\beta, R,R'}$ is the hopping matrix element between
e$_g$-orbitals $\alpha$ and $\beta$ at sites $R$ and $R'$, respectively.

\begin{figure}[tp!]
     \begin{center}
     \includegraphics[width=1.0\linewidth]{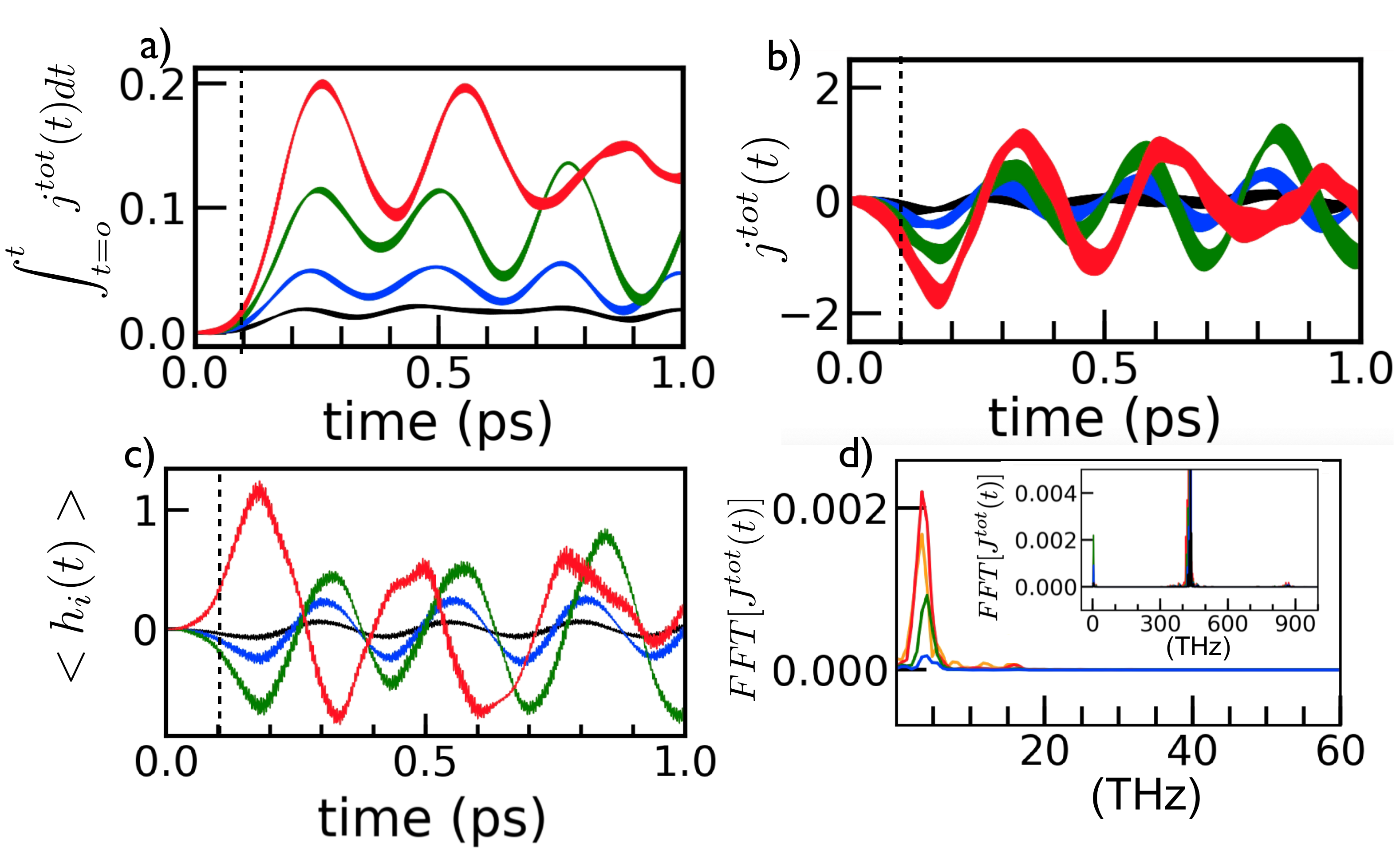}
     \caption{ a-c: Evolution of the integrated current $\int_{t=o}^tj^{tot}(t)$ (a),
     instantaneous current $j^{tot}(t)$ (b) and $h_i^{inter}(t)$ (c) in the E-type
     state at different light-intensities in the presence of atom dynamics. The dashed
     vertical line is the center of 100-fs Gaussian light-pulse. d: Fourier transform
     of $j^{tot} (t)$. The inset shows a Fourier transform of $j^{tot} (t)$ showing
     peaks in the high-frequency region. The intensities are color-coded as described in Figure 2.}
    \label{fig:fig4}
     \end{center}
\end{figure}

Figure \ref{fig:fig4} (a-b) shows the evolution of the
integrated $\int^t_{t=o} \vec{j}^{tot}(t) dt$ and instantaneous
current $\vec{j}^{tot}(t)$ at different intensities, 
both with and without atom dynamics (details available in the SI). 
In each case, the generation of a transient dc-current between 0-0.20 fs is followed by a
long-lived THz oscillatory current. This THz current
oscillations in the E-type state persist even after the light pulse,
hinting at an underlying mechanism other than the shift-current~\cite{Baltz1981,Tan2016}.
The lack of a strong frequency dependence on the intensity of the light
field (Figs.~\ref{fig:fig2}, \ref{fig:fig3}) rules out the possibility of THz current
driven by Rabi oscillations. Instead, we attribute this THz current to the collective
orbital excitations which are nonresonantly populated by the light pulse. This is
supported by the Fourier transform of $\vec{j}^{tot}(t)$ indicating a peak at $\sim 5.0$ THz,
similar to that of the orbital polarization oscillations as measured by $\vec{\tilde{P}}_{R} (t)$.
Furthermore, a real population of orbitons persists beyond the duration of the light pulse. 

To gain further insights into 
the origin of this THz-current, we look at the evolution of the asymmetry in the bonding
between the Mn site and its next-nearest (NN) neighbors on either side
along the FM zig-zag chain with similar orbital polarization. This asymmetry in the 
bonding between NN neighboring Mn-sites with similar
orbital polarization is defined by the quantity
\begin{eqnarray}
\tilde{h}^{inter}_R(t)=\sum_{\sigma}\tilde{\rho}_{\sigma,a,R,\sigma,a,{R'_+}}(t)-\tilde{\rho}_{\sigma,a,R,\sigma,a,{R'_-}}(t) 
\label{eq:bond_strenth}	 		 
\end{eqnarray}
which is computed from the off-diagonal elements $\tilde{\rho}_{\sigma,a,R,\sigma,a,R'}$
of the density matrix $\hat{\tilde{\rho}}$. The 
index $a$ in Eqn. \ref{eq:bond_strenth} indicates occupied orbital $|\Theta_l\rangle$. 
The sites $R$ and $R'_{\pm}$ represent spin-aligned 
next nearest-neighboring (NNN) pairs within individual FM zig-zag chains. 
The subscripts $+$ and $-$ indicate the NNN in the forward and backward
directions relative to $R^{th}$ Mn. Figure \ref{fig:fig4}-c shows
the dynamics of $\tilde{h}_R^{inter}(t)$. The $\tilde{h}^{inter}_R(t)$ exhibit THz oscillations
similar to $\vec{j}^{tot}(t)$. The quantity $\tilde{h}_R^{inter}(t)$ represents the
asymmetric bonding between the NNN Mn-sites.  $\tilde{h}_R^{inter}(t)$ vanishes
 for the A-type order and is non-zero for the E-type order. Indeed, the A-type order does not display
  a THz photocurrent (see SI) even though it does support orbital mode oscillations (Fig.~\ref{fig:fig3}). 

From these results we infer that in materials that support orbiton collective excitations, they can be
generically excited by optical pumping, as we have used a general model without any fine-tuning
of pump frequency, strength, or polarization. However, this condition alone is not sufficient for
observing THz currents generated by orbitons, which further requires that the orbitons be polar.
In ferroelectric materials, such as the E-type order considered here, orbitons will naturally be
polar. This phenomenon is analogous to the generation of THz phonons by optical pumps,
except in the space of orbital polarization instead of ionic polarization~\cite{merlin-generating-1997,martins-generation-2022}.  

The experimental detection of orbitons is challenging as these quasiparticles
can couple to other excitations such as phonons or magnons. These couplings make it
difficult to disentangle the contribution of orbitons in experimental measurements.
Optical pump/THz-emission spectroscopy~\cite{pettine-ultrafast-2023}, which is commonly
used to study photocurrents in strongly correlated materials, can be an alternative to
the present RIXS and Raman-spectroscopy for the experimental detection of these
quasiparticles in optically excited ferroelectrics. 

\section{Summary and conclusion}
In conclusion, our study predicts the generation of collective orbital modes
on photoexcitation in manganites exhibiting long-range orbital-order. The presence
of weak ferroelectricity ensures the manifestation of collective modes in the non-linear 
photocurrent that results in characteristic THz-emissions. These THz-emissions are
further enhanced by the dynamical modulation of intersite orbital interactions induced
by optical phonon dynamics. Complex quantum materials with strong
correlations and interactions are known to display several interesting and
non-trivial optical properties. Our work highlights that quasi-particle
excitations and their dynamics can be traced by studying the non-linear optical
responses of such materials under light illumination.

\section{Acknowledgment}
This work was primarily supported by the U.S. Department of Energy,
Office of Science, Basic Energy Sciences, Materials Sciences, and
Engineering Division, as part of the Computational Materials Sciences
Program. Part of this work was performed under the auspices of the U.S.
Department of Energy by Lawrence Livermore National Laboratory under
Contract DE-AC52-07NA27344. Additional support for data interpretation
was provided by the Molecular Foundry, a DOE Office of Science User Facility supported
by the Office of Science of the U.S. Department of Energy under
Contract No. DE-AC02-05CH11231. This research used resources of the
National Energy Research Scientific Computing Center, a DOE Office of
Science User Facility supported by the Office of Science of the U.S.
Department of Energy under Contract No. DE-AC02-05CH11231. 
This work is funded by the Deutsche Forschungsgemeinschaft 
(DFG, German Research Foundation) 217133147/SFB 1073, projects B02, B03, and C02. 
\bibliography{ref}

\begin{thebibliography}{25}%
\makeatletter
\providecommand \@ifxundefined [1]{%
 \@ifx{#1\undefined}
}%
\providecommand \@ifnum [1]{%
 \ifnum #1\expandafter \@firstoftwo
 \else \expandafter \@secondoftwo
 \fi
}%
\providecommand \@ifx [1]{%
 \ifx #1\expandafter \@firstoftwo
 \else \expandafter \@secondoftwo
 \fi
}%
\providecommand \natexlab [1]{#1}%
\providecommand \enquote  [1]{``#1''}%
\providecommand \bibnamefont  [1]{#1}%
\providecommand \bibfnamefont [1]{#1}%
\providecommand \citenamefont [1]{#1}%
\providecommand \href@noop [0]{\@secondoftwo}%
\providecommand \href [0]{\begingroup \@sanitize@url \@href}%
\providecommand \@href[1]{\@@startlink{#1}\@@href}%
\providecommand \@@href[1]{\endgroup#1\@@endlink}%
\providecommand \@sanitize@url [0]{\catcode `\\12\catcode `\$12\catcode
  `\&12\catcode `\#12\catcode `\^12\catcode `\_12\catcode `\%12\relax}%
\providecommand \@@startlink[1]{}%
\providecommand \@@endlink[0]{}%
\providecommand \url  [0]{\begingroup\@sanitize@url \@url }%
\providecommand \@url [1]{\endgroup\@href {#1}{\urlprefix }}%
\providecommand \urlprefix  [0]{URL }%
\providecommand \Eprint [0]{\href }%
\providecommand \doibase [0]{http://dx.doi.org/}%
\providecommand \selectlanguage [0]{\@gobble}%
\providecommand \bibinfo  [0]{\@secondoftwo}%
\providecommand \bibfield  [0]{\@secondoftwo}%
\providecommand \translation [1]{[#1]}%
\providecommand \BibitemOpen [0]{}%
\providecommand \bibitemStop [0]{}%
\providecommand \bibitemNoStop [0]{.\EOS\space}%
\providecommand \EOS [0]{\spacefactor3000\relax}%
\providecommand \BibitemShut  [1]{\csname bibitem#1\endcsname}%
\let\auto@bib@innerbib\@empty
\bibitem [{\citenamefont {{Reticcioli}}\ \emph {et~al.}(2019)\citenamefont
  {{Reticcioli}}, \citenamefont {{Diebold}}, \citenamefont {{Kresse}},\ and\
  \citenamefont {{Franchini}}}]{Reticcioli2019}%
  \BibitemOpen
  \bibfield  {author} {\bibinfo {author} {\bibfnamefont {M.}~\bibnamefont
  {{Reticcioli}}}, \bibinfo {author} {\bibfnamefont {U.}~\bibnamefont
  {{Diebold}}}, \bibinfo {author} {\bibfnamefont {G.}~\bibnamefont {{Kresse}}},
  \ and\ \bibinfo {author} {\bibfnamefont {C.}~\bibnamefont {{Franchini}}},\
  }\href@noop {} {\bibfield  {journal} {\bibinfo  {journal} {arXiv e-prints}\
  ,\ \bibinfo {eid} {arXiv:1902.04183}} (\bibinfo {year} {2019})},\ \Eprint
  {http://arxiv.org/abs/1902.04183} {arXiv:1902.04183 [cond-mat.mtrl-sci]}
  \BibitemShut {NoStop}%
\bibitem [{\citenamefont {Essenberger}\ \emph {et~al.}(2011)\citenamefont
  {Essenberger}, \citenamefont {Sharma}, \citenamefont {Dewhurst},
  \citenamefont {Bersier}, \citenamefont {Cricchio}, \citenamefont
  {Nordstr\"om},\ and\ \citenamefont {Gross}}]{Essenberger2011}%
  \BibitemOpen
  \bibfield  {author} {\bibinfo {author} {\bibfnamefont {F.}~\bibnamefont
  {Essenberger}}, \bibinfo {author} {\bibfnamefont {S.}~\bibnamefont {Sharma}},
  \bibinfo {author} {\bibfnamefont {J.~K.}\ \bibnamefont {Dewhurst}}, \bibinfo
  {author} {\bibfnamefont {C.}~\bibnamefont {Bersier}}, \bibinfo {author}
  {\bibfnamefont {F.}~\bibnamefont {Cricchio}}, \bibinfo {author}
  {\bibfnamefont {L.}~\bibnamefont {Nordstr\"om}}, \ and\ \bibinfo {author}
  {\bibfnamefont {E.~K.~U.}\ \bibnamefont {Gross}},\ }\href {\doibase
  10.1103/PhysRevB.84.174425} {\bibfield  {journal} {\bibinfo  {journal} {Phys.
  Rev. B}\ }\textbf {\bibinfo {volume} {84}},\ \bibinfo {pages} {174425}
  (\bibinfo {year} {2011})}\BibitemShut {NoStop}%
\bibitem [{\citenamefont {Fischer}\ \emph {et~al.}(2009)\citenamefont
  {Fischer}, \citenamefont {D\"ane}, \citenamefont {Ernst}, \citenamefont
  {Bruno}, \citenamefont {L\"uders}, \citenamefont {Szotek}, \citenamefont
  {Temmerman},\ and\ \citenamefont {Hergert}}]{Fischer2009}%
  \BibitemOpen
  \bibfield  {author} {\bibinfo {author} {\bibfnamefont {G.}~\bibnamefont
  {Fischer}}, \bibinfo {author} {\bibfnamefont {M.}~\bibnamefont {D\"ane}},
  \bibinfo {author} {\bibfnamefont {A.}~\bibnamefont {Ernst}}, \bibinfo
  {author} {\bibfnamefont {P.}~\bibnamefont {Bruno}}, \bibinfo {author}
  {\bibfnamefont {M.}~\bibnamefont {L\"uders}}, \bibinfo {author}
  {\bibfnamefont {Z.}~\bibnamefont {Szotek}}, \bibinfo {author} {\bibfnamefont
  {W.}~\bibnamefont {Temmerman}}, \ and\ \bibinfo {author} {\bibfnamefont
  {W.}~\bibnamefont {Hergert}},\ }\href {\doibase 10.1103/PhysRevB.80.014408}
  {\bibfield  {journal} {\bibinfo  {journal} {Phys. Rev. B}\ }\textbf {\bibinfo
  {volume} {80}},\ \bibinfo {pages} {014408} (\bibinfo {year}
  {2009})}\BibitemShut {NoStop}%
\bibitem [{\citenamefont {van~den Brink}\ and\ \citenamefont
  {Khomskii}(2001)}]{Brink2001}%
  \BibitemOpen
  \bibfield  {author} {\bibinfo {author} {\bibfnamefont {J.}~\bibnamefont
  {van~den Brink}}\ and\ \bibinfo {author} {\bibfnamefont {D.}~\bibnamefont
  {Khomskii}},\ }\href {\doibase 10.1103/PhysRevB.63.140416} {\bibfield
  {journal} {\bibinfo  {journal} {Phys. Rev. B}\ }\textbf {\bibinfo {volume}
  {63}},\ \bibinfo {pages} {140416} (\bibinfo {year} {2001})}\BibitemShut
  {NoStop}%
\bibitem [{\citenamefont {Saitoh}\ \emph {et~al.}(2001)\citenamefont {Saitoh},
  \citenamefont {Okamoto}, \citenamefont {Takahashi}, \citenamefont {Tobe},
  \citenamefont {Yamamoto}, \citenamefont {Kimura}, \citenamefont {Ishihara},
  \citenamefont {Maekawa},\ and\ \citenamefont {Tokura}}]{Saitoh2001}%
  \BibitemOpen
  \bibfield  {author} {\bibinfo {author} {\bibfnamefont {E.}~\bibnamefont
  {Saitoh}}, \bibinfo {author} {\bibfnamefont {S.}~\bibnamefont {Okamoto}},
  \bibinfo {author} {\bibfnamefont {K.~T.}\ \bibnamefont {Takahashi}}, \bibinfo
  {author} {\bibfnamefont {K.}~\bibnamefont {Tobe}}, \bibinfo {author}
  {\bibfnamefont {K.}~\bibnamefont {Yamamoto}}, \bibinfo {author}
  {\bibfnamefont {T.}~\bibnamefont {Kimura}}, \bibinfo {author} {\bibfnamefont
  {S.}~\bibnamefont {Ishihara}}, \bibinfo {author} {\bibfnamefont
  {S.}~\bibnamefont {Maekawa}}, \ and\ \bibinfo {author} {\bibfnamefont
  {Y.}~\bibnamefont {Tokura}},\ }\href {\doibase 10.1038/35065547} {\bibfield
  {journal} {\bibinfo  {journal} {Nature}\ }\textbf {\bibinfo {volume} {410}},\
  \bibinfo {pages} {180} (\bibinfo {year} {2001})}\BibitemShut {NoStop}%
\bibitem [{\citenamefont {Jahn}\ \emph {et~al.}(1937)\citenamefont {Jahn},
  \citenamefont {Teller},\ and\ \citenamefont {Donnan}}]{Jahn1937}%
  \BibitemOpen
  \bibfield  {author} {\bibinfo {author} {\bibfnamefont {H.~A.}\ \bibnamefont
  {Jahn}}, \bibinfo {author} {\bibfnamefont {E.}~\bibnamefont {Teller}}, \ and\
  \bibinfo {author} {\bibfnamefont {F.~G.}\ \bibnamefont {Donnan}},\ }\href
  {\doibase 10.1098/rspa.1937.0142} {\bibfield  {journal} {\bibinfo  {journal}
  {Proceedings of the Royal Society of London. Series A - Mathematical and
  Physical Sciences}\ }\textbf {\bibinfo {volume} {161}},\ \bibinfo {pages}
  {220} (\bibinfo {year} {1937})}\BibitemShut {NoStop}%
\bibitem [{\citenamefont {{van den Brink}}\ \emph {et~al.}(2000)\citenamefont
  {{van den Brink}}, \citenamefont {{Horsch}},\ and\ \citenamefont
  {{Ole{\'s}}}}]{Brink2000}%
  \BibitemOpen
  \bibfield  {author} {\bibinfo {author} {\bibfnamefont {J.}~\bibnamefont {{van
  den Brink}}}, \bibinfo {author} {\bibfnamefont {P.}~\bibnamefont {{Horsch}}},
  \ and\ \bibinfo {author} {\bibfnamefont {A.~M.}\ \bibnamefont {{Ole{\'s}}}},\
  }\href {\doibase 10.1103/PhysRevLett.85.5174} {\bibfield  {journal} {\bibinfo
   {journal} {\prl}\ }\textbf {\bibinfo {volume} {85}},\ \bibinfo {pages}
  {5174} (\bibinfo {year} {2000})},\ \Eprint
  {http://arxiv.org/abs/cond-mat/0104496} {arXiv:cond-mat/0104496
  [cond-mat.str-el]} \BibitemShut {NoStop}%
\bibitem [{\citenamefont {Inami}\ \emph {et~al.}(2003)\citenamefont {Inami},
  \citenamefont {Fukuda}, \citenamefont {Mizuki}, \citenamefont {Ishihara},
  \citenamefont {Kondo}, \citenamefont {Nakao}, \citenamefont {Matsumura},
  \citenamefont {Hirota}, \citenamefont {Murakami}, \citenamefont {Maekawa},\
  and\ \citenamefont {Endoh}}]{Inami2003}%
  \BibitemOpen
  \bibfield  {author} {\bibinfo {author} {\bibfnamefont {T.}~\bibnamefont
  {Inami}}, \bibinfo {author} {\bibfnamefont {T.}~\bibnamefont {Fukuda}},
  \bibinfo {author} {\bibfnamefont {J.}~\bibnamefont {Mizuki}}, \bibinfo
  {author} {\bibfnamefont {S.}~\bibnamefont {Ishihara}}, \bibinfo {author}
  {\bibfnamefont {H.}~\bibnamefont {Kondo}}, \bibinfo {author} {\bibfnamefont
  {H.}~\bibnamefont {Nakao}}, \bibinfo {author} {\bibfnamefont
  {T.}~\bibnamefont {Matsumura}}, \bibinfo {author} {\bibfnamefont
  {K.}~\bibnamefont {Hirota}}, \bibinfo {author} {\bibfnamefont
  {Y.}~\bibnamefont {Murakami}}, \bibinfo {author} {\bibfnamefont
  {S.}~\bibnamefont {Maekawa}}, \ and\ \bibinfo {author} {\bibfnamefont
  {Y.}~\bibnamefont {Endoh}},\ }\href {\doibase 10.1103/PhysRevB.67.045108}
  {\bibfield  {journal} {\bibinfo  {journal} {Phys. Rev. B}\ }\textbf {\bibinfo
  {volume} {67}},\ \bibinfo {pages} {045108} (\bibinfo {year}
  {2003})}\BibitemShut {NoStop}%
\bibitem [{\citenamefont {Tanaka}\ \emph {et~al.}(2004)\citenamefont {Tanaka},
  \citenamefont {Baron}, \citenamefont {Kim}, \citenamefont {Thomas},
  \citenamefont {Hill}, \citenamefont {Honda}, \citenamefont {Iga},
  \citenamefont {Tsutsui}, \citenamefont {Ishikawa},\ and\ \citenamefont
  {Nelson}}]{Tanaka2004}%
  \BibitemOpen
  \bibfield  {author} {\bibinfo {author} {\bibfnamefont {Y.}~\bibnamefont
  {Tanaka}}, \bibinfo {author} {\bibfnamefont {A.~Q.~R.}\ \bibnamefont
  {Baron}}, \bibinfo {author} {\bibfnamefont {Y.-J.}\ \bibnamefont {Kim}},
  \bibinfo {author} {\bibfnamefont {K.~J.}\ \bibnamefont {Thomas}}, \bibinfo
  {author} {\bibfnamefont {J.~P.}\ \bibnamefont {Hill}}, \bibinfo {author}
  {\bibfnamefont {Z.}~\bibnamefont {Honda}}, \bibinfo {author} {\bibfnamefont
  {F.}~\bibnamefont {Iga}}, \bibinfo {author} {\bibfnamefont {S.}~\bibnamefont
  {Tsutsui}}, \bibinfo {author} {\bibfnamefont {D.}~\bibnamefont {Ishikawa}}, \
  and\ \bibinfo {author} {\bibfnamefont {C.~S.}\ \bibnamefont {Nelson}},\
  }\href {\doibase 10.1088/1367-2630/6/1/161} {\bibfield  {journal} {\bibinfo
  {journal} {New Journal of Physics}\ }\textbf {\bibinfo {volume} {6}},\
  \bibinfo {pages} {161} (\bibinfo {year} {2004})}\BibitemShut {NoStop}%
\bibitem [{\citenamefont {Polli}\ \emph {et~al.}(2007)\citenamefont {Polli},
  \citenamefont {Rini}, \citenamefont {Wall}, \citenamefont {Schoenlein},
  \citenamefont {Tomioka}, \citenamefont {Tokura}, \citenamefont {Cerullo},\
  and\ \citenamefont {Cavalleri}}]{Polli2007}%
  \BibitemOpen
  \bibfield  {author} {\bibinfo {author} {\bibfnamefont {D.}~\bibnamefont
  {Polli}}, \bibinfo {author} {\bibfnamefont {M.}~\bibnamefont {Rini}},
  \bibinfo {author} {\bibfnamefont {S.}~\bibnamefont {Wall}}, \bibinfo {author}
  {\bibfnamefont {R.~W.}\ \bibnamefont {Schoenlein}}, \bibinfo {author}
  {\bibfnamefont {Y.}~\bibnamefont {Tomioka}}, \bibinfo {author} {\bibfnamefont
  {Y.}~\bibnamefont {Tokura}}, \bibinfo {author} {\bibfnamefont
  {G.}~\bibnamefont {Cerullo}}, \ and\ \bibinfo {author} {\bibfnamefont
  {A.}~\bibnamefont {Cavalleri}},\ }\href {\doibase 10.1038/nmat1979}
  {\bibfield  {journal} {\bibinfo  {journal} {Nature Materials}\ }\textbf
  {\bibinfo {volume} {6}},\ \bibinfo {pages} {643} (\bibinfo {year}
  {2007})}\BibitemShut {NoStop}%
\bibitem [{\citenamefont {Allen}\ and\ \citenamefont
  {Perebeinos}(1999)}]{Allen1999}%
  \BibitemOpen
  \bibfield  {author} {\bibinfo {author} {\bibfnamefont {P.~B.}\ \bibnamefont
  {Allen}}\ and\ \bibinfo {author} {\bibfnamefont {V.}~\bibnamefont
  {Perebeinos}},\ }\href {\doibase 10.1103/PhysRevLett.83.4828} {\bibfield
  {journal} {\bibinfo  {journal} {Phys. Rev. Lett.}\ }\textbf {\bibinfo
  {volume} {83}},\ \bibinfo {pages} {4828} (\bibinfo {year}
  {1999})}\BibitemShut {NoStop}%
\bibitem [{\citenamefont {Schmidt}\ \emph {et~al.}(2007)\citenamefont
  {Schmidt}, \citenamefont {Gr\"uninger},\ and\ \citenamefont
  {Uhrig}}]{Schmidt2007}%
  \BibitemOpen
  \bibfield  {author} {\bibinfo {author} {\bibfnamefont {K.~P.}\ \bibnamefont
  {Schmidt}}, \bibinfo {author} {\bibfnamefont {M.}~\bibnamefont
  {Gr\"uninger}}, \ and\ \bibinfo {author} {\bibfnamefont {G.~S.}\ \bibnamefont
  {Uhrig}},\ }\href {\doibase 10.1103/PhysRevB.76.075108} {\bibfield  {journal}
  {\bibinfo  {journal} {Phys. Rev. B}\ }\textbf {\bibinfo {volume} {76}},\
  \bibinfo {pages} {075108} (\bibinfo {year} {2007})}\BibitemShut {NoStop}%
\bibitem [{\citenamefont {Sotoudeh}\ \emph {et~al.}(2017)\citenamefont
  {Sotoudeh}, \citenamefont {Rajpurohit}, \citenamefont {Bl\"ochl},
  \citenamefont {Mierwaldt}, \citenamefont {Norpoth}, \citenamefont {Roddatis},
  \citenamefont {Mildner}, \citenamefont {Kressdorf}, \citenamefont {Ifland},\
  and\ \citenamefont {Jooss}}]{Sotoudeh2017}%
  \BibitemOpen
  \bibfield  {author} {\bibinfo {author} {\bibfnamefont {M.}~\bibnamefont
  {Sotoudeh}}, \bibinfo {author} {\bibfnamefont {S.}~\bibnamefont
  {Rajpurohit}}, \bibinfo {author} {\bibfnamefont {P.}~\bibnamefont
  {Bl\"ochl}}, \bibinfo {author} {\bibfnamefont {D.}~\bibnamefont {Mierwaldt}},
  \bibinfo {author} {\bibfnamefont {J.}~\bibnamefont {Norpoth}}, \bibinfo
  {author} {\bibfnamefont {V.}~\bibnamefont {Roddatis}}, \bibinfo {author}
  {\bibfnamefont {S.}~\bibnamefont {Mildner}}, \bibinfo {author} {\bibfnamefont
  {B.}~\bibnamefont {Kressdorf}}, \bibinfo {author} {\bibfnamefont
  {B.}~\bibnamefont {Ifland}}, \ and\ \bibinfo {author} {\bibfnamefont
  {C.}~\bibnamefont {Jooss}},\ }\href {\doibase 10.1103/PhysRevB.95.235150}
  {\bibfield  {journal} {\bibinfo  {journal} {Phys. Rev. B}\ }\textbf {\bibinfo
  {volume} {95}},\ \bibinfo {pages} {235150} (\bibinfo {year}
  {2017})}\BibitemShut {NoStop}%
\bibitem [{\citenamefont {{Rajpurohit}}\ \emph {et~al.}(2020)\citenamefont
  {{Rajpurohit}}, \citenamefont {{Jooss}},\ and\ \citenamefont
  {{Bl{\"o}chl}}}]{Rajpurohit2020}%
  \BibitemOpen
  \bibfield  {author} {\bibinfo {author} {\bibfnamefont {S.}~\bibnamefont
  {{Rajpurohit}}}, \bibinfo {author} {\bibfnamefont {C.}~\bibnamefont
  {{Jooss}}}, \ and\ \bibinfo {author} {\bibfnamefont {P.~E.}\ \bibnamefont
  {{Bl{\"o}chl}}},\ }\href {\doibase 10.1103/PhysRevB.102.014302} {\bibfield
  {journal} {\bibinfo  {journal} {Phys. Rev. B}\ }\textbf {\bibinfo {volume}
  {102}},\ \bibinfo {pages} {014302} (\bibinfo {year} {2020})}\BibitemShut
  {NoStop}%
\bibitem [{\citenamefont {Rajpurohit}\ \emph {et~al.}(2020)\citenamefont
  {Rajpurohit}, \citenamefont {Tan}, \citenamefont {Jooss},\ and\ \citenamefont
  {Bl\"ochl}}]{Rajpurohit2020_2}%
  \BibitemOpen
  \bibfield  {author} {\bibinfo {author} {\bibfnamefont {S.}~\bibnamefont
  {Rajpurohit}}, \bibinfo {author} {\bibfnamefont {L.~Z.}\ \bibnamefont {Tan}},
  \bibinfo {author} {\bibfnamefont {C.}~\bibnamefont {Jooss}}, \ and\ \bibinfo
  {author} {\bibfnamefont {P.~E.}\ \bibnamefont {Bl\"ochl}},\ }\href {\doibase
  10.1103/PhysRevB.102.174430} {\bibfield  {journal} {\bibinfo  {journal}
  {Phys. Rev. B}\ }\textbf {\bibinfo {volume} {102}},\ \bibinfo {pages}
  {174430} (\bibinfo {year} {2020})}\BibitemShut {NoStop}%
\bibitem [{\citenamefont {Mochizuki}\ \emph {et~al.}(2011)\citenamefont
  {Mochizuki}, \citenamefont {Furukawa},\ and\ \citenamefont
  {Nagaosa}}]{Mochizuki2011}%
  \BibitemOpen
  \bibfield  {author} {\bibinfo {author} {\bibfnamefont {M.}~\bibnamefont
  {Mochizuki}}, \bibinfo {author} {\bibfnamefont {N.}~\bibnamefont {Furukawa}},
  \ and\ \bibinfo {author} {\bibfnamefont {N.}~\bibnamefont {Nagaosa}},\ }\href
  {\doibase 10.1103/PhysRevB.84.144409} {\bibfield  {journal} {\bibinfo
  {journal} {Phys. Rev. B}\ }\textbf {\bibinfo {volume} {84}},\ \bibinfo
  {pages} {144409} (\bibinfo {year} {2011})}\BibitemShut {NoStop}%
\bibitem [{\citenamefont {Picozzi}\ \emph {et~al.}(2008)\citenamefont
  {Picozzi}, \citenamefont {Yamauchi}, \citenamefont {Sergienko}, \citenamefont
  {Sen}, \citenamefont {Sanyal},\ and\ \citenamefont {Dagotto}}]{Picozzi2008}%
  \BibitemOpen
  \bibfield  {author} {\bibinfo {author} {\bibfnamefont {S.}~\bibnamefont
  {Picozzi}}, \bibinfo {author} {\bibfnamefont {K.}~\bibnamefont {Yamauchi}},
  \bibinfo {author} {\bibfnamefont {I.~A.}\ \bibnamefont {Sergienko}}, \bibinfo
  {author} {\bibfnamefont {C.}~\bibnamefont {Sen}}, \bibinfo {author}
  {\bibfnamefont {B.}~\bibnamefont {Sanyal}}, \ and\ \bibinfo {author}
  {\bibfnamefont {E.}~\bibnamefont {Dagotto}},\ }\href {\doibase
  10.1088/0953-8984/20/43/434208} {\bibfield  {journal} {\bibinfo  {journal}
  {Journal of Physics: Condensed Matter}\ }\textbf {\bibinfo {volume} {20}},\
  \bibinfo {pages} {434208} (\bibinfo {year} {2008})}\BibitemShut {NoStop}%
\bibitem [{\citenamefont {Peierls}(1933)}]{Peierls1933}%
  \BibitemOpen
  \bibfield  {author} {\bibinfo {author} {\bibfnamefont {R.}~\bibnamefont
  {Peierls}},\ }\href {\doibase 10.1007/BF01342591} {\bibfield  {journal}
  {\bibinfo  {journal} {Zeitschrift f{\"{u}}r Physik}\ }\textbf {\bibinfo
  {volume} {80}},\ \bibinfo {pages} {763} (\bibinfo {year} {1933})}\BibitemShut
  {NoStop}%
\bibitem [{sup()}]{supp}%
  \BibitemOpen
  \href@noop {} {}\bibinfo {howpublished} {See Supplemental Material
  [\url{URL_will_be_inserted_by_publisher}] for more details about the the
  tight-bining model.}\BibitemShut {Stop}%
\bibitem [{\citenamefont {Kanamori}(1960)}]{Kanamori1960}%
  \BibitemOpen
  \bibfield  {author} {\bibinfo {author} {\bibfnamefont {J.}~\bibnamefont
  {Kanamori}},\ }\href {\doibase 10.1063/1.1984590} {\bibfield  {journal}
  {\bibinfo  {journal} {Journal of Applied Physics}\ }\textbf {\bibinfo
  {volume} {31}},\ \bibinfo {pages} {S14} (\bibinfo {year} {1960})},\ \bibinfo
  {note} {number: 5 Reporter: Journal of Applied Physics}\BibitemShut {NoStop}%
\bibitem [{\citenamefont {von Baltz}\ and\ \citenamefont
  {Kraut}(1981)}]{Baltz1981}%
  \BibitemOpen
  \bibfield  {author} {\bibinfo {author} {\bibfnamefont {R.}~\bibnamefont {von
  Baltz}}\ and\ \bibinfo {author} {\bibfnamefont {W.}~\bibnamefont {Kraut}},\
  }\href {\doibase 10.1103/PhysRevB.23.5590} {\bibfield  {journal} {\bibinfo
  {journal} {Phys. Rev. B}\ }\textbf {\bibinfo {volume} {23}},\ \bibinfo
  {pages} {5590} (\bibinfo {year} {1981})}\BibitemShut {NoStop}%
\bibitem [{\citenamefont {Tan}\ \emph {et~al.}(2016)\citenamefont {Tan},
  \citenamefont {Zheng}, \citenamefont {Young}, \citenamefont {Wang},
  \citenamefont {Liu},\ and\ \citenamefont {Rappe}}]{Tan2016}%
  \BibitemOpen
  \bibfield  {author} {\bibinfo {author} {\bibfnamefont {L.~Z.}\ \bibnamefont
  {Tan}}, \bibinfo {author} {\bibfnamefont {F.}~\bibnamefont {Zheng}}, \bibinfo
  {author} {\bibfnamefont {S.~M.}\ \bibnamefont {Young}}, \bibinfo {author}
  {\bibfnamefont {F.}~\bibnamefont {Wang}}, \bibinfo {author} {\bibfnamefont
  {S.}~\bibnamefont {Liu}}, \ and\ \bibinfo {author} {\bibfnamefont {A.~M.}\
  \bibnamefont {Rappe}},\ }\href {\doibase 10.1038/npjcompumats.2016.26}
  {\bibfield  {journal} {\bibinfo  {journal} {npj Computational Materials}\
  }\textbf {\bibinfo {volume} {2}},\ \bibinfo {pages} {16026} (\bibinfo {year}
  {2016})}\BibitemShut {NoStop}%
\bibitem [{\citenamefont {Merlin}(1997)}]{merlin-generating-1997}%
  \BibitemOpen
  \bibfield  {author} {\bibinfo {author} {\bibfnamefont {R.}~\bibnamefont
  {Merlin}},\ }\href {\doibase 10.1016/S0038-1098(96)00721-1} {\bibfield
  {journal} {\bibinfo  {journal} {Solid State Communications}\ }\bibinfo
  {series} {Highlights in {Condensed} {Matter} {Physics} and {Materials}
  {Science}},\ \textbf {\bibinfo {volume} {102}},\ \bibinfo {pages} {207}
  (\bibinfo {year} {1997})}\BibitemShut {NoStop}%
\bibitem [{\citenamefont {Martins}\ \emph {et~al.}(2022)\citenamefont
  {Martins}, \citenamefont {Petit}, \citenamefont {Brûlé}, \citenamefont
  {Billard}, \citenamefont {Hertz}, \citenamefont {Cluzel},\ and\ \citenamefont
  {Demichel}}]{martins-generation-2022}%
  \BibitemOpen
  \bibfield  {author} {\bibinfo {author} {\bibfnamefont {R.~J.}\ \bibnamefont
  {Martins}}, \bibinfo {author} {\bibfnamefont {M.}~\bibnamefont {Petit}},
  \bibinfo {author} {\bibfnamefont {Y.}~\bibnamefont {Brûlé}}, \bibinfo
  {author} {\bibfnamefont {F.}~\bibnamefont {Billard}}, \bibinfo {author}
  {\bibfnamefont {E.}~\bibnamefont {Hertz}}, \bibinfo {author} {\bibfnamefont
  {B.}~\bibnamefont {Cluzel}}, \ and\ \bibinfo {author} {\bibfnamefont
  {O.}~\bibnamefont {Demichel}},\ }\href {\doibase 10.1002/adom.202200357}
  {\bibfield  {journal} {\bibinfo  {journal} {Advanced Optical Materials}\
  }\textbf {\bibinfo {volume} {10}},\ \bibinfo {pages} {2200357} (\bibinfo
  {year} {2022})},\ \bibinfo {note} {\_eprint:
  https://onlinelibrary.wiley.com/doi/pdf/10.1002/adom.202200357}\BibitemShut
  {NoStop}%
\bibitem [{\citenamefont {Pettine}\ \emph {et~al.}(2023)\citenamefont
  {Pettine}, \citenamefont {Padmanabhan}, \citenamefont {Sirica}, \citenamefont
  {Prasankumar}, \citenamefont {Taylor},\ and\ \citenamefont
  {Chen}}]{pettine-ultrafast-2023}%
  \BibitemOpen
  \bibfield  {author} {\bibinfo {author} {\bibfnamefont {J.}~\bibnamefont
  {Pettine}}, \bibinfo {author} {\bibfnamefont {P.}~\bibnamefont
  {Padmanabhan}}, \bibinfo {author} {\bibfnamefont {N.}~\bibnamefont {Sirica}},
  \bibinfo {author} {\bibfnamefont {R.~P.}\ \bibnamefont {Prasankumar}},
  \bibinfo {author} {\bibfnamefont {A.~J.}\ \bibnamefont {Taylor}}, \ and\
  \bibinfo {author} {\bibfnamefont {H.-T.}\ \bibnamefont {Chen}},\ }\href
  {\doibase 10.1038/s41377-023-01163-w} {\bibfield  {journal} {\bibinfo
  {journal} {Light: Science \& Applications}\ }\textbf {\bibinfo {volume}
  {12}},\ \bibinfo {pages} {133} (\bibinfo {year} {2023})},\ \bibinfo {note}
  {publisher: Nature Publishing Group}\BibitemShut {NoStop}%
\end{thebibliography}%

\end{document}